\documentclass[10pt, a4paper, twocolumn]{article}
\pdfoutput=1

\usepackage[utf8]{inputenc}
\usepackage[T1]{fontenc}
\usepackage{lmodern, textcomp}
\usepackage[english]{babel}
\usepackage{csquotes}

\usepackage[twocolumn, bottom=3cm, textwidth=17.5cm, columnsep=.8cm]{geometry}

\usepackage{eurosym, xspace}
\usepackage[nottoc]{tocbibind}
\usepackage{authblk}

\usepackage[backend=bibtex8, citestyle=numeric-comp, maxbibnames=99, sorting=none,
			sortcites=true, firstinits=true]{biblatex}
\input{arxivshort.def}

\usepackage{amsmath, amsfonts, amssymb}

\usepackage{maths_min}

\usepackage[pdftex, breaklinks=true, linktocpage,
	colorlinks=true, urlcolor=blue, linkcolor=blue, citecolor=red,
	pdfcreator={PDFLaTeX}, pdfauthor={Harold Erbin}
]{hyperref}
\usepackage[all]{hypcap}
\usepackage[amsmath, thmmarks, hyperref, thref, framed]{ntheorem}

\hypersetup{
	pdftitle={Mabuchi spectrum from the minisuperspace},
}

\newcommand{\email}[1]{\thanks{\href{mailto:#1}{\nolinkurl{#1}}}}

\DeclareUnicodeCharacter{00A0}{~}

\title{\vspace{-2cm}Mabuchi spectrum from the minisuperspace}

\author[1]{Corinne de Lacroix\email{lacroix@lpt.ens.fr}}
\author[2]{Harold Erbin\email{erbin@lpthe.jussieu.fr}}
\author[2]{Eirik E. Svanes\email{svanes@lpthe.jussieu.fr}}
\affil[1]{\textsc{Cnrs}, \textsc{Lptens}, F-75231, Paris, France}
\affil[2]{Sorbonne Universités, \textsc{Upmc} Univ Paris 06, \textsc{Umr} 7589, \textsc{Lpthe}, F-75005, Paris, France}
\affil[2]{\textsc{Cnrs}, \textsc{Umr} 7589, \textsc{Lpthe}, F-75005, Paris, France}

\begin{document}

\maketitle

\begin{abstract}
It was recently shown that other functionals contribute to the effective action for the Liouville field when considering massive matter coupled to two-dimensional gravity in the conformal gauge.
The most important of these new contributions corresponds to the Mabuchi functional.
We propose a minisuperspace action that reproduces the main features of the Mabuchi action in order to describe the dynamics of the zero-mode.
We show that the associated Hamiltonian coincides with the (quantum mechanical) Liouville Hamiltonian.
As a consequence the Liouville theory and our model of the Mabuchi theory both share the same spectrum, eigenfunctions and -- in this approximation -- correlation functions.
\end{abstract}

\section{Introduction}

As a first step towards understanding four-dimensional quantum gravity, one may consider studying two-dimensional gravity as a toy model since many computations can be carried out exactly.
Since the partition function contains an integral over all metrics, this is equivalent to the study of statistical models on random geometries.
As such it displays rich connections with other fields of physics (statistical models, string theory) and mathematics (probability theory, random matrices and differential geometry).

In two dimensions, diffeomorphisms can be gauge-fixed by adopting the conformal gauge
\begin{equation}
	g = \e^{2\phi} g_0
\end{equation} 
where $g_0$ is a fixed metric and $\phi$ is called the Liouville mode (any quantity given in the metric $g_0$ has an index $0$).
The description of the problem is simplified since the gravity dynamics is captured by a single scalar field.
The latter is described by an effective action $S_{\text{grav}}[g_0, \phi]$ that arises from quantum effects and which provides dynamics to the metric which otherwise is non-dynamical at the classical level.
The cosmological constant already present in the classical action contributes as
\begin{equation}
	\label{quant:action:cosmological}
	S_\mu[g_0, \phi] = \mu \int \dd^2 \sigma \sqrt{g_0} \e^{2\phi}.
\end{equation} 

In the simple case when the matter theory is described by a conformal field theory (CFT), the effective action can be obtained by integrating the conformal anomaly and it is proportional to the Liouville action
\begin{equation}
	\label{quant:action:liouville}
	S_L[g_0, \phi] = \frac{1}{4\pi} \int \dd^2 \sigma \sqrt{g_0} \left( g_0^{\mu\nu} \pd_\mu \phi \pd_\nu \phi + R_0 \phi \right).
\end{equation} 
Since its introduction by Polyakov~\cite{Polyakov:1981:QuantumGeometryBosonic}, the Liouville theory has been widely studied: important steps in its definition have been the computations of the critical exponents, the computations of the spectrum and the definition of physical operators, and finally establishing that it defines a consistent CFT.
The reader is referred to the reviews~\cite{Teschner:2001:LiouvilleTheoryRevisited, Nakayama:2004:LiouvilleFieldTheory, Ribault:2014:ConformalFieldTheory} for more details.

It was shown recently in~\cite{Ferrari:2011:RandomGeometryQuantum, Ferrari:2012:GravitationalActionsTwo} that other functionals contribute to the effective action when the matter is massive
\begin{equation}
	S_{\text{grav}} = \frac{c}{6}\, S_L + \beta^2\, S_M + \cdots
\end{equation} 
Each of the functionals comes together with its coupling constant that can be computed from the parameters in the classical action and from the Liouville central charge $c = 26 - c_m$ where $c_m$ is the matter central charge.
In this letter we ignore the terms denoted by the dots and focus on the Mabuchi action $S_M[g_0, \phi]$.
For example it arises as the leading term in a small mass expansion in the case where the matter is a massive scalar field with a coupling to the curvature~\cite{Ferrari:2012:GravitationalActionsTwo}.
This action has been largely studied in differential geometry, starting with the seminal work~\cite{Mabuchi:1986:KenergyMapsIntegrating}, but it did not appear in physics until recently~\cite{Ferrari:2011:RandomGeometryQuantum, Ferrari:2012:GravitationalActionsTwo}.

Despite the fact that non-conformal matter is more relevant for describing the four-dimensional world, this topic has been mostly ignored in the $2d$ gravity literature (see~\cite{Zamolodchikov:2002:ScalingLeeYangModel, Zamolodchikov:2005:PerturbedConformalField, Zamolodchikov:2006:MassiveMajoranaFermion} for some exceptions).
For this reason it is primordial to study in more details $2d$ gravity with massive matter and to reproduce the analysis that has been done for the Liouville theory.
In particular the $1$-loop correction to the KPZ relation~\cite{Knizhnik:1988:FractalStructure2dQuantum} due to the Mabuchi functional have been computed in~\cite{Bilal:2014:2DQuantumGravity}.
Moreover, a recent series of work put forward that the very same functionals that constitute $S_{\text{grav}}$ also appear in the context of the fractional quantum Hall effect, as the leading terms in the the development of the generating functional in the number of flux quanta~\cite{Ferrari:2014:FQHECurvedBackgrounds, Can:2014:FractionalQuantumHall, Can:2015:GeometryQuantumHall}.
Hence understanding the physical properties of these functionals would shed light on this phenomena.

In this letter we propose a $(1+0)$-dimensional action that reproduces the main features of the Mabuchi action in order to describe the quantum mechanics of the Liouville zero-mode for the Mabuchi theory.
We find that the Hamiltonian of this model is equal to the (minisuperspace) Liouville Hamiltonian.
As a consequence both theories share the same spectrum and -- in this approximation -- the $2$-point and $3$-point functions are identical.
Additional considerations on the Mabuchi theory (including the derivation of our model) and the coupling of massive matter to $2d$ gravity will be presented elsewhere~\cite{deLacroix:2015:DegreesFreedom2d}.

\section{Mabuchi action}

Defining the area
\begin{equation}
	A = \int \dd^2 \sigma \sqrt{g}, \qquad
	A_0 = \int \dd^2 \sigma \sqrt{g_0},
\end{equation} 
the Mabuchi function is more conveniently expressed when the Liouville field is parametrized by the Kähler potential $K$ and the area~\cite{Ferrari:2011:RandomGeometryQuantum, Ferrari:2012:GravitationalActionsTwo}
\begin{equation}
	\label{2dgrav:eq:relation-phi-K}
	\e^{2\phi} = \frac{A}{A_0} \left(1 + \frac{A_0}{2 \pi \chi}\, \lap_0 K \right)
\end{equation}
\footnotetext{Note that our conventions are different from those in~\cite{Ferrari:2011:RandomGeometryQuantum, Ferrari:2012:GravitationalActionsTwo}: in this letter the action has been multiplied by $\pi\chi$, the Kähler potential has been divided by $\pi\chi$ and the Laplacian differs by a minus sign.
In their parametrization \eqref{2dgrav:eq:relation-phi-K} is well-defined even for $\chi = 0$.}
The Liouville field $\phi$ is uniquely determined by the pair $(A, K)$ and positivity of the exponential implies the inequality
\begin{equation}
	\lap_0 K > - \frac{2 \pi \chi}{A_0}.
\end{equation} 
In terms of functional integration, one needs to work with the partition function at fixed area: metric variations are restricted to the ones that preserve the area and the cosmological constant \eqref{quant:action:cosmological} does not contribute.
The full partition function can be recovered from the Laplace transform which amounts to introducing the cosmological constant \eqref{quant:action:cosmological} and effectively replace the area $A$ by the cosmological constant $\mu$.

In this parametrization, the Mabuchi action reads (in Lorentzian signature)~\cite{Ferrari:2012:GravitationalActionsTwo}
\begin{multline}
	\label{quant:action:mabuchi}
		S_M = - \frac{1}{4\pi} \int \dd^2 \sigma \sqrt{g_0}\; \bigg[
				- g_0^{\mu\nu} \pd_\mu K \pd_\nu K \\
				+ \left( \frac{4\pi \chi}{A_0} - R_0 \right) K
				+ \frac{4\pi \chi}{A}\, \phi \e^{2\phi}
			\bigg].
\end{multline}
Since this functional is bounded from below and convex its (Euclidean) functional integral is well-defined.

Solutions to the equation of motion correspond to constant curvature metric
\begin{equation}
	\label{2dgrav:eom:mabuchi}
	R = \frac{4\pi \chi}{A}.
\end{equation} 
The latter is identical to the equation of motion of the Liouville action \eqref{quant:action:liouville}.
At variable area this equation becomes (upon adding the cosmological constant term \eqref{quant:action:cosmological})
\begin{equation}
	\label{2dgrav:eom:mabuchi-mu}
	R = - 8 \pi \mu
\end{equation} 
which suggests the replacement
\begin{equation}
	\label{2dgrav:eq:relation-mu-A}
	\frac{\chi}{A} = - 2 \mu,
\end{equation} 
for switching between fixed and variable area.
This relation also follows by integrating \eqref{2dgrav:eom:mabuchi-mu} over the manifold and from the Legendre transformation of the cosmological constant~\cite{deLacroix:2015:DegreesFreedom2d}.

\section{Minisuperspace analysis}

The minisuperspace approximation truncates the Liouville field to its time-dependent zero-mode on flat Lorentzian space.
It is well-suited for determining the spectrum and the associated operators.
Indeed the Hilbert space can be constructed from the knowledge of what happens at one spatial point: adding space dynamics just provides multiparticle states (i.e.\ the Fock space) that are built from this Hilbert space that sits at every point.
For example the energy levels of a free scalar field are determined by the point particle approximation, which is just the quantum harmonic oscillator.

Let's consider time-dependent fields on flat spacetime
\begin{equation}
	\phi = \phi(t), \qquad
	K = K(t), \qquad
	g_0 = \eta.
\end{equation} 
The spatial dimension is compactified into a circle.
We propose the following minisuperspace action for the Mabuchi theory at variable area (in Lorentzian signature)
\begin{equation}
	\label{mini:action:mabuchi}
	S_M = - \frac{1}{2} \int \dd t \left[
			\dot K^2
			- \ddot K \ln \left( \frac{\ddot K}{4\pi \mu} \right)
			+ \ddot K
		\right]
\end{equation}
along with the following relation between $\phi$ and $K$
\begin{equation}
	\label{mini:eq:relation-phi-K}
	\e^{2\phi} = \frac{\ddot K}{4\pi\mu}.
\end{equation}

Since the Mabuchi action at variable area is not known, it is not clear how to perform rigorously the Wick rotation in order to obtain the Lorentzian action at variable area, from which the Hamiltonian formalism is sensible.
In analogy with the minisuperspace of the Liouville action, the action \eqref{mini:action:mabuchi} reproduces for the zero-mode $K(t)$ the main features of the full Mabuchi action \eqref{quant:action:mabuchi}: it contains a kinetic term for the Kähler potential and the potential term is proportional to $\phi \e^{2\phi}$.
Moreover the linear term in $K$ is not present since it vanishes for $R_0 = \cst$ and the area is replaced by the cosmological constant $\mu$ through the Laplace transform of the path integral.
For these reasons, even if the action \eqref{mini:action:mabuchi} does not correspond exactly to the minisuperspace of \eqref{quant:action:mabuchi}, it is expected that it captures the main features of the dynamics of the zero-mode and that it can be used to determine the spectrum.
Nevertheless the action \eqref{mini:action:mabuchi} can be derived in different ways under (different) mild assumptions:\footnotemark{} a detailed explanation of these various possibilities is outside the scope of this letter and we refer the reader to the companion paper~\cite{deLacroix:2015:DegreesFreedom2d}.
\footnotetext{The simplest one consists in taking the limit $R_0 \to 0$ and $A_0 \to \infty$ such that $\chi = \cst$ (and keeping $A = \cst$).
The Laplace transform of the resulting Hamiltonian is equivalent to the replacement \eqref{2dgrav:eq:relation-mu-A} and yields \eqref{mini:eq:hamiltonian-mabuchi}.
Other methods include using the Ostrogradski formalism or the fact that the kinetic and potential terms of the Mabuchi action are respectively given by the Legendre transformation of the Liouville kinetic term and of the cosmological constant action~\cite{Klevtsov:2011:2DGravityKahler}.}

As a consistency check it is straightforward to verify that the variation of \eqref{mini:action:mabuchi} agrees with the minisuperspace approximation of \eqref{2dgrav:eom:mabuchi-mu}
\begin{equation}
	\label{mini:eom:mabuchi}
	\ddot \phi = - 4 \pi \mu \e^{2\phi}.
\end{equation} 

The second-order derivatives in the action \eqref{mini:action:mabuchi} cannot be removed by integration by parts.
Fortunately the action does not depend on $K$, and the field redefinition
\begin{equation}
	J \equiv \dot K
\end{equation} 
brings \eqref{mini:action:mabuchi} to an action which is first-order in time
\begin{equation}
	\label{mini:action:mabuchi-J}
	S_M = - \frac{1}{2} \int \dd t\, \left[
			J^2
			- \dot J \ln \left(\frac{\dot J}{4\pi \mu} \right)
			+ \dot J
		\right].
\end{equation} 

The canonical momenta $P$ associated to $J$
\begin{equation}
	P = \frac{\var S_M}{\var \dot J}
		= \frac{1}{2} \ln \left( \frac{\dot J}{4\pi \mu} \right)
\end{equation}
is seen to correspond to the Liouville mode $\phi$ by comparing the previous equation with \eqref{mini:eq:relation-phi-K} written in terms of $J$.
It is well-known that a canonical transformation can be performed in order to exchange the position and momentum
\begin{equation}
	P = \phi, \qquad
	J = - \Pi,
\end{equation} 
where $\Pi$ is the conjugate momentum of the Liouville field.
In terms of these variable the Mabuchi Hamiltonian reads
\begin{equation}
	\label{mini:eq:hamiltonian-mabuchi}
	H_M = \frac{\Pi^2}{2} + 2 \pi\mu \e^{2\phi}.
\end{equation} 
It is straightforward to check that it is equivalent to the Hamiltonian of Liouville theory \eqref{quant:action:liouville} in the minisuperspace approximation.

\section{Quantization}

Since Mabuchi and Liouville theories have identical Hamiltonians, they also share the same spectrum.
For comprehensiveness we recall the canonical quantization of the Hamiltonian \eqref{mini:eq:hamiltonian-mabuchi}.

The eigenvalue equation for the Hamiltonian
\begin{equation}
	H_M \psi_p = 2 p^2\, \psi_p
\end{equation}
reduces, upon the replacement
\begin{equation}
	\Pi \longrightarrow -i\, \frac{\dd}{\dd\phi},
\end{equation} 
to the differential equation
\begin{equation}
	\label{mini:eq:schrodinger-mabuchi}
	\left(- \frac{1}{2}\, \frac{\dd^2}{\dd \phi^2}
		+ 2 \hat\mu \e^{2\phi} - 2 p^2 \right) \psi_p(\phi)
	= 0,
\end{equation} 
with the definition $\hat\mu = \pi\mu$.
The latter corresponds to the modified Bessel equation and the solutions that are well-behaved as $\phi \to \infty$ are
\begin{subequations}
\begin{align}
	\label{mini:eq:wave-function-mabuchi}
	\psi_p(\phi) &= \frac{2 \hat\mu^{- i p}}{\Gamma(- 2i p)}\; K_{2i p}(2 \sqrt{\hat\mu}\, \e^\phi) \\
		&\sim_{-\infty} \e^{2i p \phi} + R_0(p) \e^{-2i p \phi}.
\end{align}
\end{subequations} 
The eigenfunctions have been normalized such that the incoming plane wave has a unit coefficient.
The development for $\phi \to -\infty$ indicates that the waves are reflected by the potential with a reflection coefficient
\begin{equation}
	R_0(p) = \frac{\Gamma(2i p)}{\Gamma(-2i p)}\; \hat\mu^{- 2i p}.
\end{equation} 
As a consequence wave functions with $\pm p$ form a superposition and are not independent, as can be seen from the relation
\begin{equation}
	\psi_{-p}(\phi) = R_0(-p)\, \psi_p(\phi).
\end{equation} 
This divides the number of states by two.

Wave functions are normalizable only for $p \in \R$, and they form an orthogonal set
\begin{equation}
	\int_{-\infty}^\infty \dd\phi\; \conj{\psi_p}(\phi) \psi_{p'}(\phi)
		= \pi\, \dirac(p - p').
\end{equation} 
Hence physical states are associated to the eigenvalues $p \in \R_+$ and to the wave functions \eqref{mini:eq:wave-function-mabuchi}.
Moreover it can be seen that the reflection coefficient is a pure phase for these states, indicating that the potential is totally reflecting.

Finally, a semi-classical approximation of the correlation functions can be computed from integrals involving the wave functions \eqref{mini:eq:wave-function-mabuchi}.
In particular, the $3$-point function in the minisuperspace approximation is~\cite{Schomerus:2003:RollingTachyonsLiouville}
\begin{subequations}
\begin{align}
	C_0(p_1, p_2, p_3) &=
		\int_{-\infty}^\infty \dd\phi\; \psi_{p_1}(\phi) \e^{2 i p_2 \phi} \psi_{p_3}(\phi) \\
		&= \hat\mu^{- 2 \tilde p}\,
			\Gamma(2 \tilde p)
			\prod_i \frac{\Gamma\big((-1)^i 2 \tilde p_i\big)}{\Gamma(2 p_i)}
\end{align}
\end{subequations}
where we defined
\begin{equation}
	2 \tilde p = \sum_i p_i, \qquad
	\tilde p_i = \tilde p - p_i, \qquad
	i = 1, 2, 3.
\end{equation}

\section{Conclusion}

The main result of this letter is the computation of the spectrum of the Mabuchi theory.
We have shown that it coincides with the spectrum of Liouville theory.
This fact is striking since both actions have a very different origin and their forms differ vastly beyond the minisuperspace approximation (in particular the Mabuchi action is non-local in terms of the Liouville field).

On the other hand, it is not known whether the Mabuchi action defines a CFT but arguments from consistency of $2d$ gravity in the conformal gauge indicate that it should not be.
Indeed the sum of the gravity and matter actions should define a CFT of vanishing central charge in terms of the metric $g_0$: since the massive matter is not invariant its transformation needs to be compensated by the gravitational sector, which would not be invariant by itself as a consequence.
It would be very intriguing to have two theories with the same spectrum, but one being a CFT and not the other.
From the previous comments one may fear that the Liouville and Mabuchi actions describe a unique theory in two different but equivalent languages since they share many properties.
This is certainly not the case because they do not contribute in the same way to the string susceptibility exponent~\cite{Bilal:2014:2DQuantumGravity}.
This question deserves more investigation.

Obtaining a variable area formulation of the Mabuchi action is crucial in order to provide a rigorous proof of the minisuperspace action \eqref{mini:action:mabuchi}.
Moreover such a formulation would be useful for addressing other problems since it is more intuitive.

The Mabuchi theory is a key element for understanding two-dimensional quantum gravity with non-conformal matter, and for this reason it is important to study better its physical properties.
Furthermore, this would also be important in the context of condensed matter and it may even provide connections to differential geometry.

\section*{Acknowledgments}

We would like to thank Costas Bachas, Adel Bilal, Atish Dabholkar, Benoît Douçot, Frank Ferrari, Semyon Klevtsov, Raoul Santachiara and Jean-Bernard Zuber for useful discussions.
We also wish to thank Ashoke Sen for discussions and comments about the last version of the draft.

The work of E.E.S., made within the \textsc{Labex Ilp} (reference \textsc{Anr–10–Labx–63}), was supported by French state funds managed by the \emph{Agence nationale de la recherche}, as part of the programme \emph{Investissements d'avenir} under the reference \textsc{Anr–11–Idex–0004–02}.

\printbibliography[heading=bibintoc]

\end{document}